\documentstyle[12pt,definitions]{article}
\begin{document}
\begin{center}

\thispagestyle{empty}

{\normalsize\begin{flushright}
 DIAS-STP/98-03\\
ECM-UB-PF-98/03\\
FSUJ-TPI-01/98\\
Imperial/TP/97-98/22\\[10ex] \end{flushright}}
{\Large \bf  Flow Equations for Yang-Mills Theories \\ \medskip
in General Axial Gauges}\\[6ex]

{\large Daniel F. Litim\footnote{E-Mail: Litim@tholos.ecm.ub.es}${}^{,a}$ 
and Jan M. Pawlowski\footnote{E-Mail:  jmp@stp.dias.ie}${}^{,b}$}\\[5ex]

{

${}^a$Departament ECM \& IFAE,
Facultat de F\'{\i}sica\\ Univ. de Barcelona,
Diagonal 647,
E-08028 Barcelona, Spain.\\[2ex]
${}^b$Dublin Institute for Advanced Studies,\\ 
10 Burlington Road, Dublin 4, Ireland.}
\\[8ex]
 
\abstract{We present a formulation of non-Abelian gauge theories in general 
axial 
gauges using a Wilsonian (or 'Exact') Renormalisation Group. 
No 'spurious' propagator divergencies are encountered in contrast to standard 
perturbation theory.  Modified Ward identities, compatible with the flow 
equation, ensure gauge invariance of physical Green functions. The axial 
gauge $n A=0$ is shown to be 
a fixed point under the flow equation.  Possible non-perturbative 
approximation schemes  and  further applications are outlined.}

\end{center}
\newpage
\pagestyle{plain}
\setcounter{page}{1}



{\bf 1.} General axial gauges \cite{kummer}-\cite{axialregulation} 
have enjoyed considerable attention amongst the non-covariant gauges, 
especially for computations in QCD at vanishing
\cite{axialbook},\cite{qcd} 
or non-vanishing temperature \cite{axialbook},\cite{qcdT}. The main reason 
for their popularity stems from the fact that 
the ghost 
sector decouples. The number of Feynman diagrams in a
perturbative loop expansion is reduced, leading to an important
simplification from a technical point of view. Furthermore the
problem of possible Gribov copies \cite{gribovcopy}, 
generically present in covariant
gauges, is absent \cite{axialbook}. The price to pay is that the
(perturbative) propagator receives 'spurious' poles, which have to
be dealt with separately. The question about how to regularise the
propagator as to allow for a consistent loop expansion stimulated
extensive investigations \cite{axialregulation}. The intricacies
concerning these regularisations partly spoil the advantage of
having fewer diagrams to calculate. Nevertheless it has been an 
appealing gauge to e.g. calculate expectation values of Wilson loops 
which serve as order parameters for confinement. In the strong coupling 
limit they are expected to fulfil Wilson's area law which is correlated 
to a linear quark potential. The proper calculation of these expectation 
values may also 
necessitate the inclusion of topologically non-trivial configurations 
like instantons 
\cite{instaxial}. Wilson loop calculations have 
also been used as a testing ground 
for the consistency of calculations in general axial gauges. 
        
Thus it would be interesting to see
whether an alternative approach (other than standard perturbation
theory) to gauge theories in general axial gauges is available which
preserves the above mentioned benefits without encountering
'spurious' divergencies. 

In this Letter, we shall argue that these 'spurious' poles are
indeed an artifact of a perturbative loop expansion. The remedy we
propose is known as the 'Exact' (or Wilsonian) Renormalisation Group
\cite{Wilson}-\cite{average}.  This approach has already been
applied to scalar \cite{scalar}, Abelian Higgs theories 
\cite{abelianhiggs}-\cite{abelianhiggs3d} and non-Abelian gauge 
theories \cite{nonabelian}-\cite{Marco} in covariant gauges, and 
is particularly useful in
cases where perturbative expansion parameters tend to be large.
The key difference to a perturbative loop expansion consists in the
fact that flow equations integrate-out quantum fluctuations mode
by mode while integrating over infinitesimal momentum shells. This 
connects continuously the classical with the full quantum
effective action and can be interpreted as a 'coarse-graining' of
the microscopic field theory. The perturbative loop expansion lacks 
the notion of 'coarse-graining' as all modes will
be integrated-out in one step within a given loop order.
      
For the issues addressed in this Letter fermions will act only as 
spectators. Thus for the sake of simplicity we concentrate on 
the pure non-Abelian gauge theories.


{\bf 2.} Let us shortly review the appearance of 'spurious' propagator poles 
related to an axial gauge fixing in standard perturbation theory.
We will start with the Euclidean action for a pure non-Abelian
gauge theory, given in $d$ dimensions by \beq S_A[A]=\01{4}\int d^dx\
F_{\mu\nu}^a F_{\mu\nu}^a \eeq with the field strength tensor 
\beq
F_{\mu\nu}^a=\partial_\mu A^a_\nu-\partial_\nu A^a_\mu + g f^a_{\ \ 
  bc} A^b_\mu A^c_\_\nu 
\eeq 
and the covariant derivative 
\beq
D^{ab}_\mu(A)=\delta^{ab}\partial_\mu + g f^{acb}A^c_\mu, \ \ 
[t^b,t^c]={f_a}^{bc}t^a.  
\eeq 
A general axial gauge fixing for the (fixed) Lorentz vector $n_\mu$ is given by
\beq S_{\rm gf}=\01{2}\int d^dx \ n_\mu A^a_\mu\ \01{\xi n^2}\ n_\nu A^a_\nu.  
\eeq 
The gauge fixing parameter $\xi$ has mass dimension $-2$ and may also be
momentum dependent. In particular, the case $\xi=0$ ($\xi p^2 =-1$) is known
as the axial (planar) gauge. The propagator $P_{\mu\nu}$ related to
$S=S_A+S_{\rm gf}$ is  
\beq \label{Prop1}
P_{\mu\nu}=\frac{\delta_{\mu\nu}}{p^2}+\frac{n^2(1+\xi p^2)}{
  (n p)^2}\frac{p_\mu p_\nu}{p^2}
-\frac{1}{p^2}\frac{\left(n_\mu p_\nu +n_\nu p_\mu\right)}{ n
  p}.  
\eeq 
It displays the usual IR poles proportional to $1/p^2$.
In addition, we observe additional divergencies for momenta
orthogonal to $n_\mu$. These poles appear explicitly up to second order 
in $1/np$ and can even be of higher order for certain $np$-dependent 
choices of $\xi$ \cite{kummer2}.
For the planar gauge, the spurious divergencies appear only up to
first order. 

This artifact makes the application of perturbative techniques very
cumbersome as an additional regularisation for these spurious
singularities has to be introduced.


{\bf 3.} The key idea of the renormalisation group \`a la Wilson is the
step-by-step 'integrating-out of degrees of freedom'. This program
can be achieved simply by adding a scale-dependent term to the
action \cite{abelianhiggs}-\cite{Marco}, 
\beq \Delta_k S[A] = \frac{1}{2} \int \frac{d^dp}{(2\pi)^d}\ A^a_\mu
R^{ab}_{k,\mu\nu}(p) A^b_\nu.
\label{cut-off1}
\eeq 
Here we have introduced the infra-red (momentum) scale $k$
which will interpolate from some UV scale $\Lambda$ to the IR limit
$k=0$. Eq.~(\ref{cut-off1}) is quadratic in the fields and leads
therefore to a modification of the propagator.

The scale dependent
Schwinger functional $W_k[J]$ related to $S_k=S_A+S_{\rm
  gf}+\Delta_k S$ is given by 
\beq \label{Schwingerk} \exp W_k[J] =
\int {\cal D}A \exp\left\{-S_k[A]+ \int d^dx\ A^a_\mu J^a_\mu\right\}, \eeq 
and the 
scale dependent effective action $\Gamma_k$ is defined as the Legendre
transform of (\ref{Schwingerk}) 
\beq \Gamma_k[A] = \int d^dx\ J^a_\mu A^a_\mu -W_k[J]-\Delta_k
S[A],\quad A^a_\mu = \frac{\delta W_k[J]}{\delta J^a_\mu}.
\label{Gkdef}
\eeq 
For later convenience, we have subtracted $\Delta_k S$ from the Legendre
transform of $W_k$. The regulator $R_k$ enjoys the following
properties:

\begin{itemize}

\item[(i)] It has a non-vanishing limit for $p^2 \to 0$, typically
  $R\to k^2$. This precisely ensures the IR finiteness of the
  propagator at non-vanishing $k$ even for vanishing momentum $p$.

\item[(ii)] It vanishes in the limit $k\to 0$. In this limit, any
  dependence on $R_k$ drops out and $\Gamma_{k\to 0}$ reduces to
  the full quantum effective action $\Gamma$. 

\item[(iii)] For
  $k\to \infty$ (or $k\to \Lambda$ with $\Lambda$ being some UV
  scale much larger than the relevant physical scales), $R_k$
  diverges like $\Lambda^2$. Thus, the saddle point approximation
  to (\ref{Schwingerk}) becomes exact and $\Gamma_{k\to\Lambda}$
  reduces to the gauge-fixed classical action $S_A+S_{\rm gf}$.
\end{itemize}
As a consequence, the functional $\Gamma_k$ interpolates between the
gauge-fixed classical and the full quantum effective action. The
corresponding flow equation follows from (\ref{Gkdef}) as
\beq\label{flow} 
\partial_t\Gamma_k=\frac{1}{2}{\rm Tr}
\left\{\left(\Gamma_k^{(2)}+R_k\right)^{-1} \frac{\partial
    R_k}{\partial t}\right\} 
\eeq 
where the trace sums over all
momenta and indices, $t=\ln k$, and
\beqa\label{prop}
\Gamma_{k,\mu\nu}^{(2)ab}(x,x')=\di \frac{\delta^2\Gamma_k}{\delta
  A_\mu^a(x)\delta A_\nu^b(x')}.  
\eeqa 
Note that for any given
scale $k$, the main contributions to the running of $\Gamma_k$ in
(\ref{flow}) come from momenta around $p^2\approx k^2$. This
is so because $\partial_t R_k$ is peaked around $p^2\approx k^2$,
and (exponentially) suppressed elsewhere. The physics behind this is 
that a change of $\Gamma_k$ due to a further coarse graining (i.e. the
integrating-out of a thin momentum shell around $k$) is dominated
by the fluctuations with momenta around $k$. Contributions from 
fluctuations with momenta much smaller/larger than $k$ should be negligible.

{\bf 4.} What have we gained with the propagator $P_{k,\mu\nu}$ 
related to $S_k$? Let us specify the regulator as 
\beq
R^{ab}_{k,\mu\nu}(p)=\delta^{ab}\left[r(y)p^2\delta_{\mu\nu}-\tilde
  r(y) p_\mu p_\nu\right]
\label{cut-off2}
\eeq
and $y=p^2/k^2$.  A typical class of regulator functions with the
above properties is given by $(m\ge 1)$
\beq
r_m(y)=\frac{1}{\exp(y^m)-1}.
\eeq
(The limit $m\to\infty$ corresponds to the sharp cut-off limit 
\cite{Wilson}.) The
propagator takes the form 
\beq
P_{k,\mu\nu}=a_1\frac{\delta_{\mu\nu}}{p^2}+a_2 \frac{p_\mu 
p_\nu}{p^4} +a_3
\frac{n_\mu p_\nu +n_\nu p_\mu}{p^2 (np)}+a_4 \frac{n_\mu
  n_\nu}{n^2 p^2},
\label{prop_k}
\eeq with the dimensionless coefficients \beqa && \di a_1 =
1/(1+r), \quad\quad a_2 = (1+\tilde r)(1 + \xi p^2 (1+r))/z \\[1ex]
\di && a_3 = -(1+\tilde r) s^2/z,\ a_4 = -(r-\tilde r)/z \eeqa and
\bea &&s^2=(np)^2/(n^2 p^2)\\ &&z=(1+r)[(1+\tilde r) s^2 +(r-\tilde
r)(1+p^2\xi (1+r))].  \eea The most convenient choice for practical
purposes is $\tilde r=0$. The cut-off function $r$ obeys the limits
($y=p^2/k^2$) \beqa \di \lim_{y\rightarrow 0}r_m = y^{-m}, &\ \ \ \ \ 
&\di \lim_{y\rightarrow \infty}r_m = 0.  \eeqa For $\tilde r=0$ the
propagator $P_{k,\mu\nu}$ has the limits \beqa \di
\lim_{k^2\rightarrow 0}P_{k,\mu\nu} = P_{\mu\nu},\quad & & \di\ 
\lim_{p^2\rightarrow 0}P_{k,\mu\nu} = \frac{1}{k^2}\left(
  \delta_{\mu\nu}+\frac{n_\mu n_\nu}{n^2}
  \frac{1}{1+\xi k^2}\right)\delta_{m1},\\[2.5ex]
\di \lim_{k^2\rightarrow \infty}P_{k,\mu\nu} = 0\quad,\quad & & \di
\lim_{p^2\rightarrow \infty}P_{k,\mu\nu}= P_{\mu\nu}, 
\eeqa 
with
$P_{\mu\nu}$ given by (\ref{Prop1}). The infrared regulator does not
contribute for both large momenta or $k\rightarrow 0$.  For
$k\rightarrow \infty$ the propagator vanishes as all quantum modes are
suppressed. By construction, the propagator (\ref{prop_k}) is IR
finite for any $k>0$. The important observation is now the following:
In contrast to the perturbative propagator $P_{\mu\nu}$, the limit of
$P_{k,\mu\nu}$ for $np\to 0$ is finite! This holds true even
 for an arbitrary choice of $\xi(p,n)$ and leads to 
\beq
P_{k,\mu\nu}  = \di \frac{1}{1+r}\frac{\delta_{\mu\nu}}{p^2}+
\frac{1+\tilde
  r}{(1+r)(r-\tilde r)}\frac{p_\mu p_\nu}{p^4}
-\di\frac{1}{(1+r)(1+p^2\xi (1+r))} \frac{n_\mu n_\nu}{n^2
  p^2}\label{prop_np}.  
\eeq
Thus (\ref{prop_np}) is perfectly
well-behaved and finite for all momenta $p$ (as long as the regulators
$r$ and $\tilde r$ have not been chosen to be identical). It is noteworthy 
that the 'spurious' divergencies are absent as soon as the infra-red 
behaviour of the propagator is under control.


{\bf 5.} Now we shall discuss the implications of gauge invariance, 
and in particular the question about how to ensure gauge
invariance of physical Green functions. To that end we consider
the Schwinger functional as given in (\ref{Schwingerk}). Invariance
of the measure ${\cal D} A$ under the infinitesimal gauge transformation 
$A\rightarrow A+ D\alpha$ 
leads to the so-called modified Ward-identity (mWI) 
\beqa
&&\di 0=\left\langle D^{ab}_\mu J^b_\mu - 
n_\mu D_\mu^{ab}(x)\  \frac{1}{n^2 \xi}\ n_\mu A_\mu^b(x)
- D_\mu^{ab}(x)\ R_{k,\mu\nu}^{bc}\  A^c_\nu(x) \right\rangle_J.
\label{WI} \eeqa
The mWI contains quadratic powers of the gauge field, which is a
consequence of the gauge fixing and the cut-off term. Usually one
converts the WI into a linear form by introducing ghosts and using
BRST-invariance. However, due to the cut-off term this is no longer
possible. With 
\beq\label{G_k}
G_{k,\nu\mu}^{ab}(x,x')=\left(\Gamma_{k,\nu\mu}^{(2)ab}+
R^{ab}_{k,\nu\mu}
\right)^{-1}(x,x')
\eeq 
the mWI can be converted into 
\beqa \di {\cal
  W}_k^a[A]&\equiv& \di D_\mu^{ab}(x)\frac{\delta
 \Gamma_k[A]}{\delta A^b_\mu(x)} \di-n_\mu D_\mu^{ab}(x)\ 
\frac{1}{n^2\xi}
n_\nu A_\nu^b(x) \\[2ex]&&\di -g \int d^dy\ f^{abc}\left(\frac{1}{n^2\xi}n_\mu
  n_\nu\delta^{cd} +R^{cd}_{k,\mu\nu}\right)(x,y)
  G_{k,\nu\mu}^{db}(y,x) = 0
\label{mWI}\eeqa
The first term is the usual covariant functional derivative of 
$\Gamma_k$, the second one stems from the gauge fixing, while the
third one, a consequence of the regularisation and a possible 
momentum dependence of $\xi$, introduces
non-local contributions. Note that 
the $R_k$-dependent term vanishes for both $R_k\to\infty$ and
$R_k\to 0$. Both (\ref{WI}) and (\ref{mWI}) reduce to the
usual Ward identity in the limit $k\to 0$ \cite{axialbook}. 
For $k\neq 0$, they explicitly 
depend on the regulator, and the relations amongst
different $n$-point functions of the theory as implied by (\ref{mWI}) will 
depend on the particular form of the coarse-graining. 
                     
We will now derive the important result which is that the  flow equation 
(\ref{flow}) and the mWI (\ref{mWI}) are compatible. To that end we 
calculate the scale dependence of (\ref{mWI}). Using the flow 
equation (\ref{flow}) for
$\Gamma_k$, one can check explicitly that 
\beq \label{compatible}
\partial_t {\cal
  W}^a_k = -\frac{1}{2}{\rm Tr}\left( G_k \frac{\partial
    R_k}{\partial t} G_k \frac{\delta}{\delta
    A}\times\frac{\delta}{\delta A}\right){\cal W}^a_k 
\eeq
where the trace sums over momenta and internal indices. 
It follows that if $\Gamma_k$ fulfils the Ward identity at some
scale $k_0$, and if it evolves according to the flow equation
(\ref{flow}), then the new action will again fulfil the
corresponding mWI.\footnote{It can be shown that (\ref{compatible}) 
is valid for general linear gauges. For covariant linear gauges, 
a similar result was obtained in \cite{GaugeInv}.}
Thus it is in principle sufficient to fulfil the mWI at some
initial scale $\Lambda$ in order to ensure gauge invariance of the 
physical Green functions, that is in the limit $k\to 0$.

Let us comment on the possible momentum dependence of $\xi$. As
$\xi$ is dimensionful, it is natural to consider it as an operator.
In perturbation theory, for example, the planar gauge $\xi(p)=-1/p^2$ 
introduces a
momentum dependence in order to reduce the degree of the spurious
divergencies. This is no longer necessary as the
divergencies are absent  in the present approach.  
Therefore we may restrict 
ourselves to a momentum-independent gauge parameter. This is an
important simplification, as the mWI (\ref{mWI}) reduces to
\beqa  
\di D_\mu^{ab}(x)\frac{\delta
  \Gamma_k[A]}{\delta A^b_\mu(x)} \di-\frac{1}{n^2\xi} 
n_\mu\partial^x_\mu\ n_\nu A^a_\nu (x)\di -g \int d^dy\ f^{abc}R^{cd}_{k,\mu\nu}
(x,y) G_{k,\nu\mu}^{db}(y,x)= 0.  
\label{mWI2}\eeqa 
As a consequence the possible tensor structure of vertices is 
considerably simplified. However, 
one may ask whether the limit $\xi\rightarrow 0$ defines the axial gauge 
also for momentum dependent $\xi(p)$. In perturbation theory, the answer is 
yes \cite{kummer2}. In the present context, it suffices 
to show that the term
\beqa\label{0} \di
\int d^dy\ f^{abc}\frac{n_\mu n_\nu}{n^2\xi (x,y)} G_{k,\nu\mu}^{cb}(y,x)
\eeqa
in (\ref{mWI}) vanishes in the limit $\xi(p)\rightarrow 0$. That this is 
indeed the case can be seen as follows: Expanding (\ref{0}) in powers of 
the gauge 
field yields expressions containing the 
effective propagator $G_k$ and $n$-point vertices $\Gamma_k^{(n>2)}$. 
Both $G^0_k$ and $\Gamma_k^{(n>2)}$ are completely 
determined via the (finite) flow equation. Thus  they cannot exhibit singular 
terms proportional to $1/\xi$. 
Using in addition $n_\mu G_{k,\mu\nu}={\cal O}(\xi)$ (due to the gauge 
fixing)  we find  that (\ref{0}) is at least of 
order ${\cal O}(\xi)$ for $\xi\rightarrow 0$, which establishes the axial 
gauge even for a momentum dependent gauge fixing.


{\bf 6.} As a first application of this formalism we shall argue that
the choice $\xi=0$ corresponds to a fixed point w.r.t. the flow
equation. In principle, this can be shown through an
explicit computation of the flow equation for $\xi(k)$. However,
this would necessitate an explicit Ansatz for $\Gamma_k$ (and would
therefore be only an approximation). In contrast, we present an
argument which makes use only of the mWI.  With the preceding result 
in mind we can safely restrict 
ourselves to the case of $\xi$ being momentum independent, although the 
result will hold true  for general $\xi(p)$. First note that
$\xi$ enters the mWI both explicitly and implicitly. The $\xi$ 
appearing explicitly corresponds to the choice of $\xi$ at some
initial scale $\Lambda$, $\xi\equiv\xi(\Lambda)$. An implicit
dependence occurs through the dependence of $\Gamma_k$ on the scale
dependent $\xi(k)$. Let us choose $\xi(\Lambda)=0$ with
$\Gamma_\Lambda$ solving (\ref{mWI}) and {\em assume} that
$\xi(k)\neq 0$ for some $k<\Lambda$. This means in particular that
$\Gamma_k$ will no longer contain a singular term $\sim 1/\xi$.
Thus the only singular term appearing in the mWI is the term
explicitly proportional to $1/\xi$. One can always find an $A^a$
such that $n\partial\ nA^a$ does not vanish. Therefore $\Gamma_k$ with
$\xi(k)\neq 0$ can not be a solution of (\ref{mWI}) for
$\xi(\Lambda)=0$. But this cannot be true as the compatibility of
the flow equation and the mWI (\ref{compatible}) implies that
$\Gamma_k$ solves the mWI. It follows that $\xi(k)=0$ for
$\xi(\Lambda)=0$. Hence the axial gauge is indeed a fixed point of the flow
equation. Note that this argument does not involve any
approximations regarding $\Gamma_k$.
     
We conclude that out of all general axial gauges (even
momentum dependent ones) the  axial ($\xi=0$) gauge is singled out
and appears to be the natural choice. Furthermore, $\xi=0$ is 
well-suited for actual computations as both the flow equation and
the mWI are rather simple in that case.


{\bf 7.} A second application concerns the question about how to control
gauge invariance for an {\em approximate} solution of the
flow equation.  Generally speaking, the task of solving the flow
equation for gauge theories faces two main problems. The first one
is to find a solution of (\ref{mWI}) at some initial scale
$k=\Lambda$ in order to ensure gauge invariance of the physical
Green function in the limit $k\to 0$. The second one concerns an
{\it Ansatz} for the functional form of $\Gamma_k[A]$. As not all
possible operators can be taken into account, the validity of
(\ref{mWI}) for $k<\Lambda$ (automatically ensured only for the 
full effective action  $\Gamma_k[A]$) can no longer be taken for 
granted and has to be checked independently.

As an example, consider the action $\Gamma_0=S_A+S_{\rm gf}$ for $\xi=0$ at the
scale $k=0$. 
Obviously, $\Gamma_0$ is a solution of  the mWI (\ref{mWI2}). The
flow equation 
can be integrated analytically in leading order of perturbation
theory, 
replacing $\Gamma_k$ through $\Gamma_0$ on the r.h.s. of 
(\ref{flow}), to give
\beqa \label{1loop}
\Gamma_k & = &\di \Gamma_0+\frac{1}{2}\Tr\ln\left( 
\Gamma^{(2)}_0+R_k\right)-
\frac{1}{2}\Tr\ln \Gamma^{(2)}_0. 
\eeqa 
The mWI without the $R_k$-dependent term is solved by $\Gamma_0$. For the 
remaining terms in (\ref{mWI2}) we obtain in leading order   
\beqa\label{wi1loop}
\di\frac{1}{2} D_\mu^{ab}(x)\frac{\delta
  (\Tr\ln( \Gamma^{(2)}_0+R_k)-\Tr\ln \Gamma^{(2)}_0 )}
{\delta A^b_\mu(x)}
\di-g\int d^dy\ f^{abc}R^{cd}_{k,\mu\nu}(x,y)
G_{k,\nu\mu}^{db}(y,x) =0 . 
\eeqa 
$\Gamma_0^{(2)}$ is the second derivative of $\Gamma_0$ 
with respect to the gauge field (\ref{prop}). 
We use the commutator 
$[D\frac{\delta}{\delta A},\frac{\delta^2}{(\delta A)^2}]$ and 
(\ref{mWI2}) for $\Gamma_0$ to obtain 
\beqa\label{expansion}
\di\frac{1}{2} D_\mu^{ab}(x)\frac{\delta
  \Tr\ln(\Gamma^{(2)}_0+R_k)}{\delta A^b_\mu(x)} =\di  
-g\int d^dy\ f^{abc}\Gamma^{(2)cd}_{0,\mu\nu}(x,y)
\left(\frac{1}{\Gamma^{(2)}_0+R_k}\right)^{db}_{\nu\mu}(y,x). 
\eeqa 
For $k=0$, (\ref{expansion}) is simply zero. Using 
$(\Gamma^{(2)}_0+R_k)^{-1}=G_{k}+{\cal O}(g)$ and 
inserting (\ref{expansion}) into (\ref{wi1loop}) results in 
\beqa\label{check}
&  &\di  
-g\int d^dy\ f^{abc}\left(\Gamma^{(2)cd}_{0,\mu\nu}
+R^{cd}_{k,\mu\nu} \right)(x,y) 
G_{k,\nu\mu}^{db}(y,x)\ \sim -g^2 f^{abc}
\delta^{bc}=0.   
\eeqa 
Thus we have shown that the compatibility of the flow and the mWI
can be maintained even within an approximate solution. It is 
straightforward to show that this holds
true systematically even for higher orders within a
perturbative loop expansion \cite{Marco}. 
       
Note that in \cite{GaugeInv} the 1-loop perturbative
compatibility of the flow equation was checked explicitly (for
covariant gauges) for the scale dependent gluon mass parameter
using a BRST-formulation. This is a rather non-trivial task since
the flow equation as computed directly from (\ref{flow}) or from
the corresponding Slavnov-Taylor identity receives contributions
from quite different diagrams (involving ghosts and gauge fields).
However, in our formulation (without ghosts) the consistency check
is rather simple and is done without problems for the entire
effective action.

In more general situations, and especially within non-perturbative
regions, it is not obvious how the compatibility between flow and mWI
 of a given truncation 
can be maintained. However, it is still possible to exploit the
compatibility condition and to use it as a control mechanism for
the Ansatz itself. A natural implementation would be to use the mWI
as a flow equation for some of the relevant couplings (like mass
terms) of the theory. Comparing the flow of these operators with
the flow as derived directly from (\ref{flow}) allows one to control
the domain of validity of a given truncation.\footnote{See
  \cite{GaugeInv}, \cite{heavy} where a similar line of reasoning has been 
employed on QCD in covariant gauges.}
   
For approximations beyond perturbation 
theory the flow equation (\ref{flow}) and the mWI (\ref{mWI2}) 
can  also be used to control the dependence on $n_\mu$ of the effective 
action and thus generalise the observations of \cite{rebhan} 
to the case with a cut-off term. 

{\bf 8.} Let us finally comment on the computation of the 1-loop 
$\beta$-function.\footnote{All the 
details of the computation shall be presented elsewhere \cite{1-loop}.} 
This is another crucial test for the 
viability of this method. In standard perturbation theory, the 'spurious' 
singularities seem to play an important r$\hat{\rm o}$le and 
do already contribute  
on the 1-loop level \cite{axialbook}. In the present context they are 
absent throughout and it would be interesting to check that the 
correct 1-loop running is still coming out. 

The running gauge coupling $g_k^2$ is related to the 
scale dependent wave function renormalisation of the field strength 
$Z_{F,k}$ via $g^2_k=k^{d-4}g^2/Z_{F,k}$. Using the Ansatz 
\beq
\Gamma_k= \014\int d^dx\ Z_{F,k}\, F_{\mu\nu}^a F_{\mu\nu}^a 
+ S_{\rm gf}[Z_{F,k}^{1/2}A]
\eeq
and \eq{flow} one can deduce the 1-loop flow for $Z_{F,k}$ by projecting on 
the $F^2$-terms in the flow equation. This is self-consistent at the 
1-loop level. The traces 
involved in \eq{flow} can be calculated using heat kernel 
methods\footnote{Here, the heat kernel is {\it not} used 
as a regularisation since \eq{flow} is finite anyhow.}  and one  obtains 
\beq\label{runZ}
\partial_t \ln Z_{F,k} = \0{11N}{24\pi^2} g^2 + {\cal O}(g^4).
\eeq 
 The $\beta$-function follows 
immediately from \eq{runZ} as  
\beq
\beta_{g^2}\equiv \partial_t g^2_k = -\0{11N}{24\pi^2}g_k^4 + {\cal O}(g_k^6),
\eeq
i.e. the well-known 1-loop result. The above (universal) result can be 
shown to hold for any regulator with the properties (i) -- (iii). 


{\bf 9.} To sum up, we 
have developed a self-contained and self-consistent formulation
of QCD in general axial gauges which allows full control
over the IR behaviour of the theory. In contrast to standard
perturbation theory the 'spurious' propagator singularities are
naturally absent.  Gauge invariance of physical Green functions is
controlled via the mWI which is shown to be compatible with the
flow equation. The absence of ghosts results in a rather simple
expression for the mWI. We have shown that of all
general axial gauges the axial gauge $nA=0$ is singled out as it
corresponds to a fixed point under the Wilsonian flow. Possible
ways of finding approximate solutions even beyond perturbation
theory have been briefly indicated. The calculation of the 
1-loop running of the gauge coupling has been outlined 
and the $\beta$-function agrees with the known perturbative result.

This formulation seems to be a good starting point for theories
where Lorentz covariance is naturally broken. This is the case for
QCD at finite temperature where the heat bath singles
out a rest frame. It is straightforward to apply the present approach 
within the imaginary time formalism. The propagator would still come 
out without any 'spurious' divergencies, in contrast to the recent 
proposal \cite{pietroni} based on a renormalisation group for 
the temperature fluctuations only.
Another possible application concerns quantum
field theories with both electrically and magnetically charged
U(1)-fields, as this necessitates the introduction of a fixed
Lorentz vector \cite{monopoles}. In either case, the
Lorentz vector $n_\mu$ should be used for the axial gauge fixing as
described above. It may also necessitate the inclusion of topologically 
non-trivial configurations into the effective average action \cite{insts}. 
We hope to report on these matters in the future.

\section*{Acknowledgments}
We wish to thank U. Ellwanger for numerous discussions, and the L.P.T.H.E.
Orsay for hospitality at an early stage of the work.
This work was
supported in part by the European Commission under the Human Capital
and Mobility program, contract number CHRX-CT94-0423.


\begin{thebibliography}{99}
\def\BOOK#1#2#3#4{#1 {\sc #2}, #3, #4}
\def\PRA#1#2#3#4#5{ #1   Phys.~Rev.~{\bf A #3} (19#4) #5}
\def\PRB#1#2#3#4#5{#1   Phys. Rev.~{\bf B #3} (19#4) #5}
\def\PRL#1#2#3#4#5{#1   Phys. Rev.~Lett.~{\bf #3} (19#4) #5}
\def\PRC#1#2#3#4#5{#1   Phys. Rev.~{\bf C #3}  (19#4) #5}
\def\PRD#1#2#3#4#5{#1   Phys. Rev.~{\bf D #3} (19#4) #5}
\def\PRE#1#2#3#4#5{#1   Phys. Rev.~{\bf E #3} (19#4) #5}
\def\PRep#1#2#3#4#5{#1   Phys. Rep.~{\bf  #3} (19#4) #5}
\def\NPB#1#2#3#4#5{#1   Nucl. Phys.~{\bf B #3} (19#4) #5}
\def\PLB#1#2#3#4#5{#1   Phys. Lett.~{\bf B #3} (19#4) #5}
\def\ibid#1#2#3#4#5{#1   {\it ibid.~}{\bf #3} (19#4) #5}
\def\PTP#1#2#3#4#5{#1   Prog. Theor.~Phys.~{\bf B #3} (19#4) #5}
\def\SSC#1#2#3#4#5{#1   Solid State Comm.~{\bf  #3} (19#4) #5}
\def\EPL#1#2#3#4#5{#1   Europhys. Lett.~{\bf #3} (19#4) #5}
\def\JCP#1#2#3#4#5{#1   J.~Phys. (Paris) {\bf  #3} (19#4) #5}
\def\JPA#1#2#3#4#5{#1   J.~Phys. {\bf A  #3} (19#4) #5}
\def\JPB#1#2#3#4#5{#1   J.~Phys. {\bf B  #3} (19#4) #5}
\def\JPC#1#2#3#4#5{#1   J.~Phys. {\bf C  #3} (19#4) #5}
\def\ZPC#1#2#3#4#5{#1   Z.~Phys. {\bf C  #3} (19#4) #5}
\def\JETP#1#2#3#4#5{#1   Soviet Physics JETP Lett.~{\bf #3} (19#4) #5}
\def\MPLA#1#2#3#4#5{#1   Mod.~Phys. Lett.~{\bf A  #3} (19#4) #5}
\def\PA#1#2#3#4#5{#1   Physica {\bf A  #3} (19#4) #5}
\def\PS#1#2#3#4#5{#1   Physics {\bf   #3} (19#4) #5}
\def\AP#1#2#3#4#5{#1   Ann. Phys. {\bf  #3} (19#4) #5}
\def\IJMPA#1#2#3#4#5{#1   Int.~J. Mod. Phys.~ {\bf A  #3} (19#4) #5}
\def\LNC#1#2#3#4#5{#1   Lett.~Nuevo Cimento {\bf   #3} (19#4) #5}
\def\PPR#1#2#3{#1    Preprint #3}
\def\and#1#2#3{{\bf #1} (19#2) #3}

\bibitem{kummer}W. Kummer, Acta Phys. Austriaca {\bf 14} (1961)
  149; {\bf 41} (1975) 315; 
R.L. Arnowitt and S.I. Fickler, Phys. Rev. {\bf 127} (1962) 1821; 
Yu.L. Dokshitzer, D.I. Dyakonov and S.I. Troyan, Phys. Rep. {\bf 58}
  (1980) 269; A.H. Mueller, {\it ibid.} {\bf 73} (1981) 238. 
\bibitem{axialbook} A. Bassetto, G. Nardelli and R. Soldati, {\it
Yang-Mills Theories in algebraic non-covariant gauges: Canonical
Quantization and Renormalization}, Singapore, World Scientific (1991);   
G. Leibbrandt, {\it Noncovariant Gauges}, Singapore, Word Scientific (1994). 
\bibitem{qcd}F.Lenz, H.W.L. Neus and M. Thies, Annals Phys. {\bf 242} (1995) 429; 
H. Reinhardt, Phys. Rev. {\bf D 55} (1997) 2331. 
\bibitem{qcdT}\PLB{K. Kajantie and J. Kapusta,}{INFRARED LIMIT OF THE
AXIAL GAUGE GLUON PROPAGATOR AT HIGH
TEMPERATURE}{110}{82}{299}; \ZPC{K.A. James,}{THE TEMPORAL AXIAL GAUGE
AT FINITE TEMPERATURE EXPLORED USING THE REAL TIME
FORMALISM}{48}{90}{169}; \PLB{M. Kreuzer and H. Nachbagauer,}{POLE
PRESCRIPTION IN AXIAL GAUGE AT FINITE TEMPERATURE}{271}{91}{155}.
\bibitem{kummer2}\PRD{D.M. Capper and G. Leibrandt}{}{25}{82}{1002}; 
W. Kummer, Phys. Rev. {\bf D 37} (1988) 454. 
\bibitem{rebhan}P. Gaigg, O. Piguet, A. Rebhan and M. Schweda,
 Phys. Lett. {\bf B 175} (1986) 53.
\bibitem{axialregulation} A. Bassetto, I. Lazzizzera and R. Soldati,
 Phys. Lett. {\bf B 107} (1981) 278; {\bf B 131} (1983) 177;
 Nucl. Phys. {\bf B 236} (1984) 319;  {\bf B 276} (1986) 517; 
H. Cheng and E.C. Tsai, Phys.~Rev.~Lett. {\bf 57} (1986) 511; 
Phys. Lett. {\bf B 176} (1986) 130; Phys.~Rev. {\bf D 36} (1987) 3196; 
J.P. Leroy, J.Micheli and G.C. Rossi, Nuovo Cim. {\bf A 84} (1984)
 270; Nucl.~Phys. {\bf B 232} (1984) 511; 
H. H\"uffel, P.V. Landshoff and J.C. Taylor, Phys.~Lett. {\bf B 217} (1989) 
147; B.J. Hand and G. Leibbrandt, {\tt hep-th/9511140}.
\bibitem{gribovcopy} V. Gribov, Nucl.~Phys.~{\bf B 139} (1978) 1.
\bibitem{instaxial}
G.C. Rossi and M. Testa, Nucl.~Phys.~{\bf B 163} (1980) 109;  {\bf B 176} 
(1980) 477;  {\bf B 237} (1984) 442; Phys.Rev. {\bf D 29} (1984) 2997.
\bibitem{Wilson}K.~G.~Wilson and I.~G.~Kogut, Phys. Rep. {\bf 12} (1974) 75;
F.~Wegner and A.~Houghton, Phys. Rev. {\bf A 8} (1973) 401.
\bibitem{ERG} J.~Polchinski, Nucl. Phys. {\bf B 231} (1984) 269;
T.~Hurd, Commun. Math. Phys. {\bf 124} (1989) 153;
G.~Keller, C.~Kopper and M.~Salmhofer, Helv. Phys. Acta {\bf 65} (1992) 32;
G.~Keller and C.~Kopper, Commun. Math. Phys. {\bf 161} (1994) 515.  
\bibitem{average}\NPB{C.~Wetterich,}{}{352}{91}{529}; 
\PLB{}{}{301}{93}{90}; \IJMPA{T.~Morris}{}{9}{94}{2411}.
\bibitem{scalar}\ZPC{C.~Wetterich,}{}{57}{93}{451}; \NPB{C.~Wetterich and 
N.~Tetradis,}{}{422}{94}{541}; \ibid{N.~Tetradis and 
D.F.~Litim,}{}{B 464}{96}{492}.
\bibitem{abelianhiggs}\NPB{M.~Reuter and C.~Wetterich,}{}{391}{93}{147}; 
\and{B 408}{93}{91}; \and{B 427}{94}{291}; F. Freire and 
C. Wetterich, Phys. Lett. {\bf B 380} (1996) 337.
\bibitem{abelianhiggs4d}\PLB{D.F.~Litim,}{}{393}{97}{103}; 
\MPLA{D.F.~Litim, C.~Wetterich and N.~Tetradis,}{}{12}{97}{2287}.
\bibitem{abelianhiggs3d}B.~Bergerhoff, F.~Freire, D.F.~Litim, S.~Lola 
and C.~Wetterich, Phys.~Rev.~{\bf B 53} (1996) 5734;  B.~Bergerhoff, 
D.F.~Litim, S.~Lola and C.~Wetterich, Int.~J.~Mod.~Phys. {\bf A 11} 
(1996) 4273.
\bibitem{nonabelian}M.~Reuter and C.~Wetterich, 
Nucl. Phys. {\bf B 417} (1994) 181; M.~Bonini, M.~D'Attanasio and 
G.~Marchesini, 
{\it ibid. } {\bf B 418} (1994) 81;  {\bf B 421} (1994) 429; 
 {\bf B 437} (1995) 163; Phys. Lett. {\bf B 346} (1995) 87. 
\bibitem{GaugeInv}U.~Ellwanger, Phys. Lett. {\bf B 335} (1994) 364.
\bibitem{heavy}   U.~Ellwanger, M.~Hirsch and A.~Weber, Z. Phys. {\bf C 69}
  (1996) 687; Eur. Phys. J. {\bf C1} (1998) 563. 
\bibitem{insts} J.~M. Pawlowski, {\tt hep-th/9605037}, to be published in Phys. Rev. {\bf D}. 
\bibitem{Marco}\PLB{M. d'Attanasio and T. Morris,}{}{378}{96}{213}.
\bibitem{pietroni}\NPB{M. d'Attanasio and M. Pietroni,}{}{498}{97}{443}.
\bibitem{monopoles}\PRD{D. Zwanziger,}{}{3}{71}{880}; \PRep{M. 
Blagojevic and P. Senjanovic,}{}{157}{88}{233}.
\bibitem{1-loop}D.F. Litim and J.M. Pawlowski, under completion.
\end{thebibliography}
\end{document}